\documentclass[12pt,thmsa]{article}
\usepackage{amsfonts}
\usepackage{graphicx}
\usepackage{epsfig}

\usepackage{graphics}

\makeatletter
\newcommand{\row}[1]%
{\mathord{\buildrel{\lower3pt%
\hbox{$\scriptscriptstyle\rightarrow$}}\over #1}}

\newcommand{\dyadic}[1]{\mathord{\dyadic@rrow{#1}}}
\newcommand{\dyadic@rrow}[1]{
\begin{picture}(12,12)(-1,0)
\put(-3,12){\makebox(0,0)[t]{$\scriptscriptstyle\downarrow$}}
\put(-3,13){\makebox(0,0)[l]{$\scriptscriptstyle\longrightarrow$}}
\put(5,0){\makebox(0,0)[b]{$#1$}}
\end{picture}
}

\newcommand{\bra}[1]{\bigl\langle #1 \bigr|}
\newcommand{\ket}[1]{\bigl| #1 \bigr\rangle}

\begin{document}

%
\begin{center}
{\Large  Teleportation of Accelerated   Information}
\\
\vspace{0.5cm}
N. Metwally\\
Math. Dept., Faculty of Science, Aswan University, Aswan,
Egypt.\\
Math. Dept., College of Science, University of Bahrain, Bahrain\\
\end{center}

 \begin{abstract}
A theoretical quantum teleportation protocal is suggested to
teleport accelerated and non-accelerated information over
different classes of accelerated quantum channels. For the
accelerated information, it is shown  that the fidelity of the
teleported state increases as the entanglement of the initial
quantum channel increases. However as the difference between the
accelerated channel and the accelerated information decreases the
fidelity increases. The fidelity of the non accelerated
information increases as the entanglement of the initial quantum
channel increases, while  the accelerations of the quantum channel
has a little effect. The possibility of sending quantum
information over accelerated quantum channels is much better than
sending classical information.

 \end{abstract}

\section{Introduction}

Dynamics of entangled qubits in non-inertial frames is
investigated in the literature, from different points of view. For
example, Unruh and Wald \cite{Unruh} have investigated the nature
of the interaction between a quantum field and accelerating
particle detector. The entanglement dynamics of Dirac fields in
non-inertial frames is  studied by Alsing et al.  \cite{ Alsing1}.
Generating entangled photons from the vacuum by accelerated
measurements has been investigated by Han et. al \cite{Mux}. In
\cite{Jieci1}, the effect of quantum decoherence generated by the
Unruh effect on the dynamics of a maximum entangled state is
discussed.  Said and Adami \cite{Said} have invistigated the
Einstein-Podolsky-Rosen correlation in Kerr-Newman space time.
Montero et al. \cite{Mon} have studied the case of families of
fermionic field states in non inertial frames and the dynamics of
multipartite entanglement of fermionic systems in non-inertial
frames is considered  by Wang and Jing \cite{Jieci}. The classical
and quantum correlations of scalar field in the inflationary
universe have been discussed by Nambu and  Ohsumi \cite{Nem}. The
analogy between static quantum emitters coupled to a single mode
of quantum field and accelerated Unruh-DeWitt detectors is shown
by Rey et al.  \cite{Mark}. Demonstrating  entanglement generation
between mode pairs of a quantum field in a single, rigid cavity
that moves nonuniformly in Minkowski space-time has been
investigated  by Friis et al. \cite{Nic}. The correlations between
the accelerated field modes and the modes in an inertial reference
cavity are discussed in \cite{Nic1}.

Implementation of quantum information tasks between accelerated
partners represents one of the most important topics in the
context of quantum communication. One of these  tasks is quantum
teleportation, which enables to send information between two users
without sending the source itself which carries this information
\cite{Bennt}. Some efforts have been achieved to perform quantum
teleportation in non-inertial frames. For example, the possibility
of performing some quantum information protocols in non-inertial
is investigated by Enk et al. \cite{Enk}. Teleportation by using
three parties with an accelerated receiver is investigated by Jin
et. al \cite{Jin}.
 Quantum teleportation in the presence of Unruh effect
is investigated by Landulfo et al. \cite{Landu}.

Recently, Metwally  \cite{Metwally1} investigated the dynamics of
a general two qubits system in non inetrial  frame analytically,
by assuming that both  of its subsystems are differently
accelerated. The most useful classes of travelling entangled
channels in non-inertial frames are classified. The conclusions
was, the maximum entangled channels and a pure state with large
degree of entanglement could be used  as a quantum channel to
perform quantum teleportation. This motivated  to suggest
 a theoretical quantum teleportation
protocol between two users share an accelerated channels. Our
protocol is different from the other in  two aspects: {\it
firstly} we assume that the subsystems of the quantum channel are
differently accelerated and two classes of useful quantum channels
for quantum teleportation are considered, namely maximum and
partial entangled channels \cite{Metwally1}. {\it Second},  the
desired teleported state is assumed to be  accelerated in the same
frame of the sender. Also, the possibility of teleporting
non-accelerated information by using accelerated channels is
discussed.

This paper is organized as follows: in Sec.2, we review the
dynamics of  quantum channels between  two users in non-inertial
frame, where  two classes of initial quantum channels are
considered: generic pure state  and maximum entangled state.  In
Sec.3, the generated entangled channels between the users are
employed as  quantum channels to perform the original quantum
teleportation protocol. Two classes are considered for the
teleported state: one is accelerated with a uniform acceleration
while the second is non accelerated. The fidelity of the
teleported state is quantified for different accelerated channels.
Finally, the results are summarized in Sec.4.

\section{Accelerated channels}
Lets assumed that  two users Alice and Rob that share a two-qubit
state, where each qubit of this state (Alice and Rob's qubit)
travels  in non-inertial frames with a uniform acceleration. In
this context, it is  assumed that the source supplies the users
with two types of initially entangled state: maximum or partial
entangled pure state.

Let the partners initially share  pure state of the form,
\begin{equation}\label{Pure}
\rho_{p}= \frac{1}{4}\left(
\begin{array}{cccc}
(1-q)&
-p&p&-(1-q)\\
\\
p&(1+q)&-(1+q)&p\\
\\
p&-(1+q)& (1+q)&-p\\
\\
-(1-q)&p& -p&(1-q)
\end{array}
\right),
\end{equation}
where $q=\sqrt{1-q^2}$.  This channel is  represented by one
parameter, $p$. For  $p=0$  one gets  a  maximum entangled state
of  Bell's type. However for larger values of $p$, the pure state
turns into a partial entangled state and completely seperable for
$p=1$ \cite{Englert,Englert2}. To investigate the teleportation by
using accelerated channels, we find the new channel in the
non-inertial frame by using the Unruh modes transformation
\cite{un,Asp},

\begin{eqnarray}\label{Un}
\ket{0}_i&=&\mathcal{C}_i\ket{0}_I\ket{0}_{II}+\mathcal{S}_i\ket{1}_I\ket{1}_{II},
\nonumber\\
\ket{1}_i&=&\ket{1}_{I}\ket{0}_{II},
\end{eqnarray}
where $\mathcal{C}_i=\cos r_i, \mathcal{S}_i=sin r_i,$ with  $tan
r_i=e^{-\pi\omega_i \frac{c}{a_i}}$, $a_i$ is the acceleration,
$\omega_i$ is the frequency of the travelling qubits, $c$ is the
speed of light and $i=A$ (Alice), $B$ (Bob). The accelerated
channel in the region $I$  is given by \cite{Metwally1},

\begin{equation}\label{PurI}
\tilde\rho_{p}= \left(
\begin{array}{cccc}
\varrho_{11}&
\varrho_{12}&\varrho_{13}&\varrho_{14}\\
\\
\varrho_{21}&\varrho_{22}&\varrho_{23}&\varrho_{24}\\
\\
\varrho_{31}&\varrho_{32}& \varrho_{33}&\varrho_{34}\\
\\
\varrho_{41}&\varrho_{42}& \varrho_{43}&\varrho_{44}
\end{array}
\right),
\end{equation}
where the elements $\varrho_{ij}$ and $i,j=1..4$ are given by,
\begin{eqnarray}
\varrho_{11}&=&\frac{1}{4}\Bigl[\frac{1-q}{4}(1+2\cos2r_1)+\frac{3-q}{4}\cos
2r_2\Bigr], \quad \varrho_{12}=-\frac{p}{8}\cos r_2
\bigl[3-cos2r_1\bigr],
\nonumber\\
\varrho_{13}&=&\frac{p}{8}\cos
r_1\bigl[1+\cos2r_2\bigr],\quad\quad\quad\quad\quad
\varrho_{14}=-\frac{1-q}{4}\cos r_1\cos r_2,
\nonumber\\
\varrho_{21}&=&-\frac{p}{8}\cos 2r_2\bigl[3-\cos 2r_1\bigr], \quad
\varrho_{22}=\frac{1}{4}\Bigl[q-1+2(q+1)\cos2 r_1+(q+3) \cos
2r_2\Bigl],
 \nonumber\\
 \varrho_{23}&=&-\frac{1+q}{4}\cos r_1\cos r_2,\quad\quad
 \varrho_{24}=\frac{p}{8}\cos r_1\bigl[3-\cos2 r_2\bigl],
 \nonumber\\
 \varrho_{31}&=&\varrho_{13}, \quad
 \varrho_{32}=\varrho_{23},
 \nonumber\\
 \varrho_{33}&=&\frac{1}{4}\Bigl[\frac{1+q}{2}\cos 2
 r_1+\frac{3+q}{4}(1+\cos 2r_2)\Bigl], \quad
 \varrho_{34}=-\frac{p}{8}\cos r_1\cos r_2(1+\cos 2 r_1),
 \nonumber\\
\varrho_{41}&=&\varrho_{14},\quad \varrho_{42}=\varrho_{24},
\nonumber\\
\varrho_{44}&=&\frac{1}{4}\Bigl[\frac{9-q}{4}-\frac{1+q}{2}\cos2
r_1-\frac{3+q}{4}\cos 2r_2\Bigl].
\end{eqnarray}

\begin{figure}[t!]
  \begin{center}
  \includegraphics[width=30pc,height=20pc]{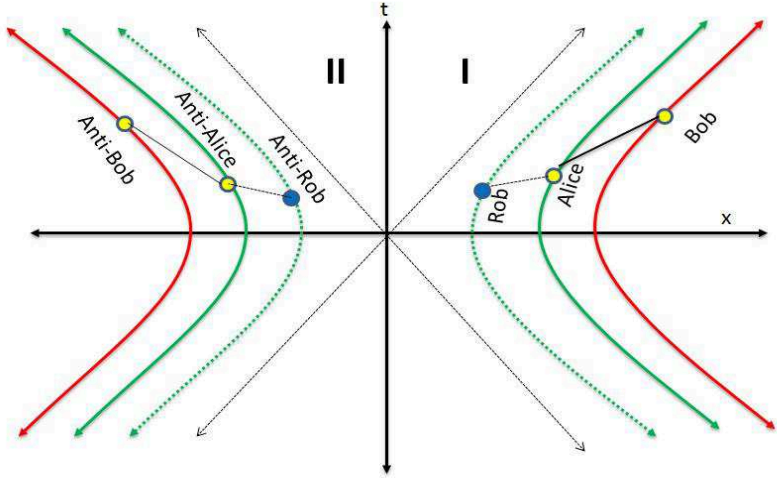}~
     \caption{Mainkowski space time diagram for Alice, Bob and Rob who has the unknown information.
   An accelerated users Alice and Bob travel  with uniform accelerations $r_1$ and $r_2$ respectively in the region I.
     Theses users are causally disconnected from their Anti-users in the region $II$.
     The unknown information is coded in Rob's
     qubit which is  accelerated with another different acceleration $r_3$.
     Quantum teleportation is achieved in the   region $I$, where Alice's aim is teleporting Rob's state to Bob.
        }
  \end{center}
\end{figure}

\section{Teleportation}

 In this section, it is assumed that the partners share the   accelerated state(\ref{PurI}) as quantum channel
to perform quantum teleportation \cite{Bennt}. A source supplies
Alice with unknown information  coded in Rob's state as:
\begin{equation}\label{u}
\ket{\psi}_R=\alpha\ket{0}+\beta\ket{1},
\end{equation}
where $|\alpha|^2+|\beta|^2=1$. The description of the suggested
protocol is described  in Fig.(1). It is assumed that Alice 's
qubit and Rob's qubit are accelerated in the same frame with
different accelerations. In this context we consider two
situations:
\begin{enumerate}
\item{ \it The teleported information is accelerated}

In this case, we assume that the coded information in the state
(\ref{u}) is accelerated according to Eq.(\ref{Un}), with an
acceleration $r_3$. The density operator  in the region I, which
carries the unknown information is given by,
\begin{equation}\label{MI}
\rho_{R}=|\alpha|^2\cos^2 r_3\ket{0}\bra{0}+|\alpha\sin
r_3+\beta|^2 \ket{1}\bra{1}.
\end{equation}
 To teleport the unknown informatin which
is coded on the state (\ref{MI}), Alice and Bob will use the
accelerated state (\ref{PurI}) in the region I as an quantum
channel. For this purpose the partners perform the original
quantum teleportation protocol \cite{Bennt} as  follows:
\begin{enumerate}
\item Alice performs CNOT gate on her qubit and the given qubit
state  (5)\cite{Cory}. After performing this operation, the final
state of the system is given by,

\begin{equation}
\rho_s=CNOT\rho_{R}\otimes \tilde\varrho_p CNOT
\end{equation}
\item Alice applies Hadamard gate on the given qubit  follows by
performing  Bell measurements on the given qubit and her own
qubit. Then,she transmits her results to Bob via classical
channel.

\item Depending on the results of Alice's measurements, Bob
performs a suitable unitary operations on his own qubit  to get
the original message.If Alice measures the state
$\ket{\phi_{+}}=\frac{1}{\sqrt{2}}(\ket{00}+\ket{11})$, then Bob
will get the state,
\begin{equation}
\rho_B=\mu_{00}\ket{0}\bra{0}+\mu_{01}\ket{0}\bra{1}+\mu_{10}\ket{1}\bra{0}+\mu_{11}\ket{1}\bra{1},
\end{equation}
where,
\begin{eqnarray}
\mu_{00}&=&\frac{1}{4}\Big((B_1+B_2)(\varrho_{11}+\varrho_{33})+(B_1-B_2)(\varrho_{13}+\varrho_{31})\Bigl),
\nonumber\\
\mu_{01}&=&\frac{1}{4}\Bigl((B_1+B_2)(\varrho_{12}+\varrho_{34})+(B_1-B_2)(\varrho_{14}+\varrho_{32})\Bigl),
\nonumber\\
\mu_{10}&=&\frac{1}{4}\Bigl((B_1+B_2)(\varrho_{21}+\varrho_{43})+(B_1-B_2)(\varrho_{23}+\varrho_{41})\Bigl),
\nonumber\\
\mu_{11}&=&\frac{1}{4}\Bigl((B_1+B_2)(\varrho_{22}+\varrho_{44})+(B_1-B_2)(\varrho_{24}+\varrho_{42})\Bigl),
\end{eqnarray}
and $B_1=|\alpha|^2\cos^2 r_3$, $B_2=|\alpha\sin r_3+\beta|^2$.
The fidelity of the teleported  information which is coded in the
state(\ref{MI}) is given by,
\begin{equation}\label{Fpa}
\mathcal{F_P}_{a}=B_1\mu_{00}+B_2\mu_{11}.
\end{equation}
\end{enumerate}
\item{\it The teleported information is non-accelerated}

 In this case the teleported information is coded in the state,
\begin{equation}\label{nonac}
\rho_{Rs}=|\alpha|^2\ket{0}\bra{0}+\alpha\beta^*\ket{0}\bra{1}+\alpha^*\beta\ket{1}\bra{0}+|\beta|^2\ket{1}\bra{1},
\end{equation}
Alice and Bob perform the original teleportation protocol as
described above. At the end of the protocol Bob will get  the
state
\begin{equation}
\rho_B=\nu_{00}\ket{0}\bra{0}+\nu_{01}\ket{0}\bra{1}+\nu_{10}\ket{1}\bra{0}+\nu_{11}\ket{1}\bra{1},
\end{equation}
where,
\begin{eqnarray}
\nu_{00}&=&\frac{1}{4}\Bigl\{\varrho_{11}+\varrho_{33}+(|\alpha|^2-|\beta|^2)(\varrho_{13}+\varrho_{31})-
\lambda^+\varrho_{12}-
\lambda^-(\varrho_{13}+\varrho_{13}-\varrho_{33}\Bigl\},
\nonumber\\
\nu_{01}&=&\frac{1}{4}\Bigl\{\varrho_{42}+\varrho_{34}+(|\alpha|^2-|\beta|^2)(\varrho_{14}+\varrho_{32})-
\lambda^+\varrho_{42}-
\lambda^{-}(\varrho_{34}+\varrho_{32}-\varrho_{14}\Bigl\},
\nonumber\\
\nu_{10}&=&\frac{1}{4}\Bigl\{\varrho_{21}+\varrho_{43}+(|\alpha|^2-|\beta|^2)(\varrho_{23}+\varrho_{41})-
\lambda^+\varrho_{21}-
\lambda^{-}(\varrho_{43}+\varrho_{41}-\varrho_{23}\Bigl\},
\nonumber\\
\nu_{11}&=&\frac{1}{4}\Bigl\{\varrho_{22}+\varrho_{44}+(|\alpha|^2-|\beta|^2)(\varrho_{24}+\varrho_{42})-
\lambda^+\varrho_{22}-
\lambda^-(\varrho_{44}+\varrho_{42}-\varrho_{24}\Bigl\},
\nonumber\\
\end{eqnarray}
with $\lambda^+=\alpha^*\beta+\alpha\beta^*$ and
$\lambda^-=\alpha\beta^*-\alpha^*\beta$. In this case the fidelity
of the teleported state is given by,
\begin{equation}
\mathcal{F_P}_{s}=|\alpha|^2\nu_{00}+\alpha\beta^*\nu_{01}+\alpha^*\beta\nu{10}+|\beta|^2\nu_{11}.
\end{equation}
\end{enumerate}

\begin{figure}
  \begin{center}
   \includegraphics[width=25pc,height=14pc]{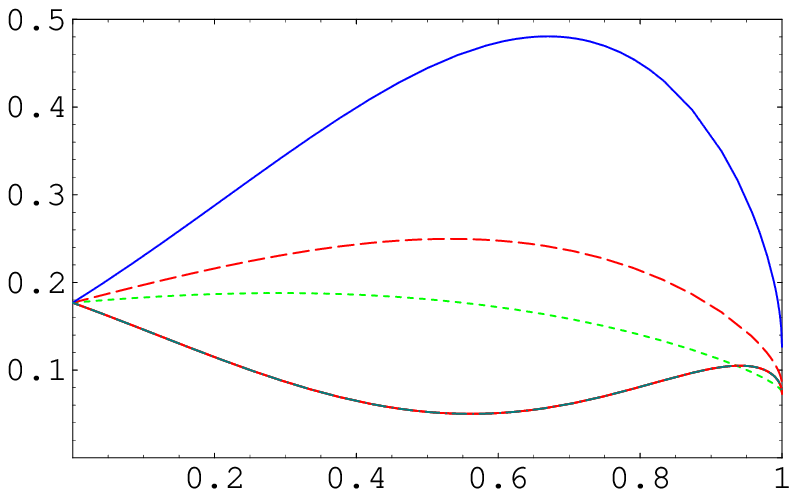}
\put(-325,90){$\mathcal{F_P}_{a,s}$}
 \put(-130,-10){$\alpha$}
      \caption{The fidelity  $\mathcal{F_P}_a$  of the  accelerated  teleportated   state, where the partners used
     a quantum channel initially prepared in maximum entangled state, i.e. $p=0$.
  The   dot, dash and solid curves are plotted for $r_3=0.1,0.3$ and $0.8$
  respectively while   i.e $r_1 =r_2=0.7$. The most lower curve
     represents the fidelity of the non accelerated information
     $\mathcal{F_P}_s$.
     }
  \end{center}
\end{figure}

The  fidelity  $\mathcal{F_P}_a$ of the accelerated teleported
state  and the non accelerated teleported state
$\mathcal{F_P}_{s}$ are described in Fig.(1). In this case, we
assume that the partners share a quantum channel  initially
prepared in maximum entangled states (MES). The two qubits  are
accelerated with an equal acceleration i.e. $r_a=r_b=0.7$, while
the teleported state is accelerated with different accelerations.
The fidelities are plotted against the parameter $\alpha$ which
represents the structure of the teleported state.  The three upper
curves represent the fidelity of the accelerated teleported state
$\mathcal{F_P}_a$, while the lower one represents the fidelity of
the non-accelerated teleported state $\mathcal{F_P}_{s}$. From
this figure, it is clear that, $\mathcal{F_P}_a$ increases as the
acceleration $r_3$ of the teleported state  increases. However the
fidelity $\mathcal{F_P}_a$ decreases if the difference between the
channel's accelerations and the teleported state's acceleration
increases. The fidelity $\mathcal{F_P}_{s}$ of the non-accelerated
teleported state( the most lower curves) decreases as $\alpha$
increases. Also, it is clear that all the fidelities coincide.
This shows that the behavior of $\mathcal{F_P}_{s}$ depends on the
entanglement of the accelerated channel.
\begin{figure}[t!]
  \begin{center}
   \includegraphics[width=25pc,height=14pc]{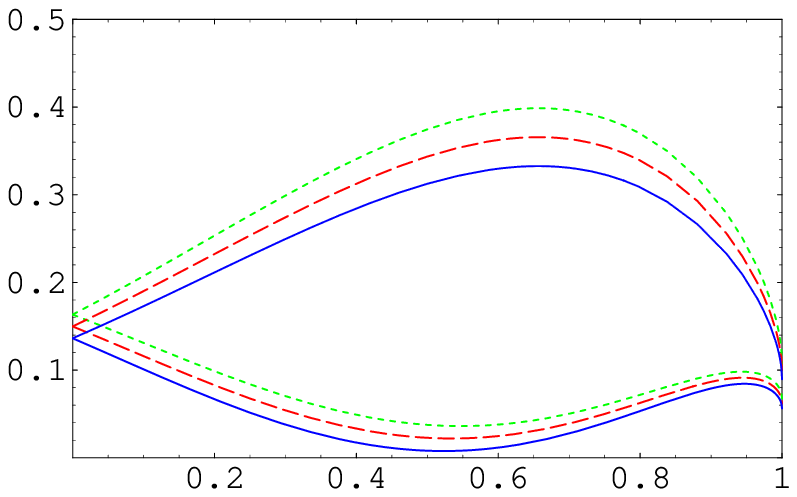}
    \put(-130,-10){$\alpha$}
    \put(-325,90){$\mathcal{F_P}_{a,s}$}
     \caption{The  same as Fig.(1) but the partners share an initially a pure
     state. The most upper three curves  represent  the fidelity of the accelerated teleported state
$\mathcal{F_P}_{a}$, while the most lower curves represent the
fidelity of non-accelerated teleported state $\mathcal{F_P}_{s}$.
The  solid, dash dot  are for $p=0.1,0.2,0.3$ respectively, where
$r_1=r_2=r_3=0.7$.}
  \end{center}
\end{figure}

\begin{figure}
  \begin{center}
  \includegraphics[width=19pc,height=14pc]{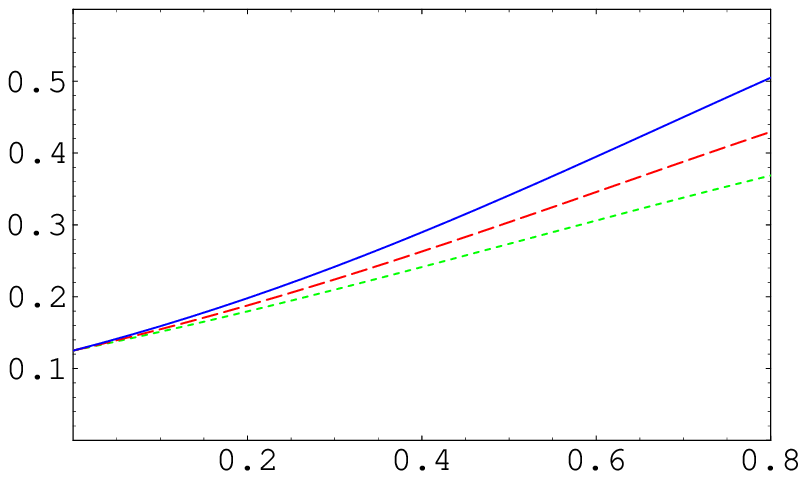}~
  \put(-100,-5){$r_3$}
  \put(-5,100){$\mathcal{F_P}_a$}
   \includegraphics[width=19pc,height=14pc]{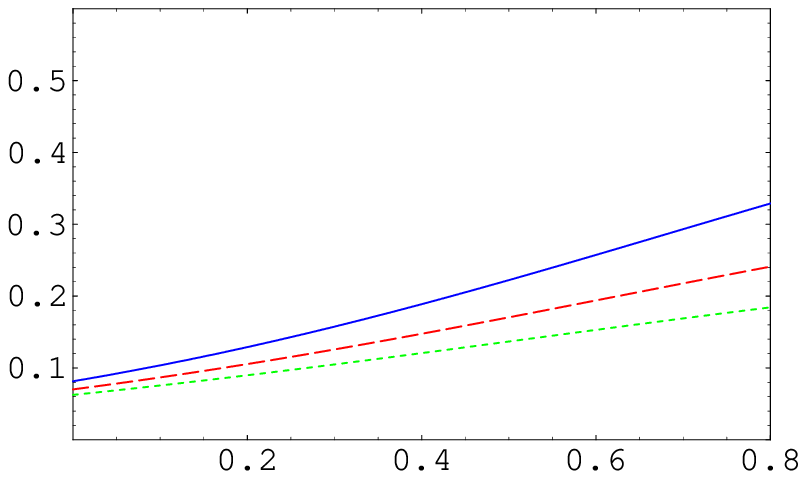}
   \put(-100,-5){$r_3$}
   \put(-480,100){$\mathcal{F_P}_a$}
     \caption{The fidelity of  the accelerated information which is initially coded
     in the state $\ket{\psi_u}=\frac{1}{\sqrt{2}}(\ket{0}+\ket{1})$, where the used accelerated channel is prepared initially in
     (a) maximum entangled state  and (b)pure state (Partial entangled
     state with $p=0.5$). The solid,dash and dot for $r_1=r_2=0.8.,
     0.5$
     and $0.0001$ respectively.}
  \end{center}
\end{figure}

In Fig.(3),  the behavior of $\mathcal{F_P}_{a}$  and
$\mathcal{F_P}_{s}$ for different entangled pure states is
displayed. The fidelity  $\mathcal{F_P}_{a}$ of the   accelerated
information behaves  similar to that depicted for MES in Fig.(2).
In this figure, it is assumed that the accelerations of the
channel and the teleported state  are equal i.e.,
$r_1=r_2=r_3=0.7$. The fidelity increases  for small values of
$p$, namely, the partners starting with entangled channels have a
large degree of entanglement. However for larger values of $p$,
the degree of entanglement of the quantum channel decreases and
consequently the fidelity of the teleported state decreases for
both accelerated and non-accelerated teleported states.

From Figs.$(2\&3)$, we  conclude  that, for the accelerated
information the fidelity $\mathcal{F_P}_{a}$ depends on the
difference between the acceleration of the quantum channel and the
acceleration of the teleported state as well as the degree of
entanglement of the quantum channel between the partners. However
for the non-accelerated teleported state, the  fidelity
$\mathcal{F_P}_{s}$, depends only on the degree of entanglement.
The accelerated maximum entangled channel teleports the
accelerated and non-accelerated information  with larger fidelity
compared with that teleported by partial entangled channels.

 The behavior of the teleported information depends on
the parameter $\alpha$, where the initial  information is coded
with  a probability $\alpha^2$ in the state $\ket{0}$ and with
probability $\beta=\sqrt{1-\alpha^2}$ in the state $\ket{1}$. For
$\alpha$ or $\beta$ equals $"1"$, this is equivalent to coding
classical information, while for $0<\alpha<1$ and $0<\beta<1$,
this corresponds to coding quantum information. It is clear that,
teleportating  quantum informatiom is much better than
teleportating  classical information
 over an accelerated channel.

 Fig.4, displays the behavior of the fidelity $\mathcal{F_P}_a$
for different accelerated channels, for $|\alpha|^2=0.5$. In
Fig.(4a), we assume that the partners share initially a maximum
entangled state. For larger values of $r_1$ and $r_2$, the
fidelity $\mathcal{F_P}_a$ increases as $r_3$ increases. However,
 the fidelity of the teleported state decreases as the
acceleration of the teleported state decreases. As shown in
Fig.(4b), starting from a partial entangled channel between the
partners, the fidelity of the teleported state is smaller than
that depicted for its corresponding one  in Fig.(4a), where the
partners share a MES. Also at $r_3=0$, the fidelity of the
teleported state $\mathcal{F_P}_a$ depends on the degree of the
quantum channel between the partners. This explain why the initial
fidelities  at $r_3=0$ have different values.

\section{Conclusion}
Quantum teleportation  in non-inertial frames, is investigated by
using  different classes of accelerated channels between the
users: maximum and pure entangled channels. It is shown that,
these accelerated quantum channels  represent  useful classes for
quantum teleportation.  In this protocol we consider the desired
information to be coded either in an accelerated  or
non-accelerated state. Our results show that the fidelity of the
teleported state depends on the initial degree of entanglement of
the used quantum channel, the accelerations of the quantum
channel, the acceleration of  teleported state and the structure
of the coded information.

Performing the original quantum teleportation protocol by using a
maximum  entangled channel is much better than using a partial
entangled state. It is shown  that, the fidelities of the
accelerated and non accelerated teleported state   using an
accelerated channel prepared in a maximum entangled state is much
larger  than that depicted for accelerated partial entangled
channels. The fidelity of the non-accelerated state depends only
on the initial entanglement of the accelerated quantum channels
between the two users.

 On the other
hand, the fidelity of the teleported information increases when
the difference between the channel's acceleration and the
teleported state's acceleration is small. The possibility of
sending classical information  via quantum accelerated channel is
much better than teleporting quantum information.

{\it In conclusion},  one can  teleport information between two
moving users sharing  a maximum or partial accelerated entangled
channels. If we can control  the accelerations (for channel and
teleported state) to be the same, the accelerated information can
 then be  teleported with high fidelity. Sending classical
information via accelerated channel is much better than sending
quantum information by using same  channels.

\end{document}